\documentstyle[twocolumn]{article}
\title{Comment on ``Linguistic Features of Noncoding DNA Sequences''}
\author{N. E. Israeloff, M. Kagalenko, and K. Chan\\
        \small{israeloff@nuhub.neu.edu, mkagalen@lynx.dac.neu.edu,
kechan@lynx.dac.neu.edu}\\
        Dept. of Physics\\
        Northeastern University\\
        Boston MA 02115}
\date{January 26 1995}
\begin{document}
\maketitle
In a recent letter~\cite{ref:mantegna}, Mantegna et.\ al.\ report that
certain statistical signatures of natural language can be found in
non-coding DNA sequences.  The vast majority of DNA in higher
organisms including humans consists of non-coding sequences whose
function, if any, is unknown.  Hence this new analysis is quite
important.  It suggests, as the authors concluded, ``the possible
existence of one  (or more than one) structured biological language(s)
present in non-coding DNA sequence''.  Previous work from this group
and others also showed that DNA sequences have long-range power-law
correlations~\cite{ref:peng}~\cite{ref:voss}, which are found in
non-coding regions but not in coding regions~\cite{ref:peng}.

Since disorder or randomness dominates many natural phenomena which exhibit
power-law correlations it is reasonable to ask whether such
correlations alone can produce the statistical features attributed to
the presence of language. Here we show that random noise with
power-law correlations, similar to the ubiquitous ``1/f'' noise,
exhibits the same ``linguistic'' statistical signatures reported in
ref.~1 for non-coding DNA. We conclude that these signatures by
themselves cannot distinguish language from noise.

As in ref.~1 we
carried out the Zipf analysis of ``word'' frequency  vs.
rank as well as the Shannon analysis of redundancy.  Noise with
spectral density of the form $S(f)\sim f^{-\beta}$ was analyzed.  The exponent
$\beta$ can be related to the  exponent $\alpha$ used in the DNA walk
correlation
analysis by $\beta=2\alpha-1$.~\cite{ref:peng}~\cite{ref:comment} Noise was
synthesized by numerically
filtering white-noise with a power-law filtering function.  The white
noise was derived from either a gaussian random number generator or
from amplified thermal noise from a 1 M$\Omega$ resistor, with no difference
in outcome.  We also analyzed $1/f$ noise from a  Josephson junction,
which had the same statistical behavior as synthesized
noise.~\cite{ref:tobe} Signals were
binned into four amplitude ranges with equal weight and assigned
values $0-3$ so as to provide a  four letter ``alphabet'', like that of
DNA.  This method is equivalent to sampling with a 2-bit
analog-to-digital converter.   The analysis consisted of sampling
contiguous blocks of length $n$  (an $n$-tuple).  The $n$-point
sampling window was sequentially shifted by one point until the entire sequence
was sampled.  The number of occurrences of each such $n$-tuple was
counted, then
the $n$-tuples were ranked from highest to lowest frequency of
occurrence.

The Zipf plot of word frequency vs. rank has power-law
behavior with exponent $\zeta=-1$ for natural languages.  For
non-coding DNA $\zeta$ ranged from $0.289$ to
$0.537$.~\cite{ref:mantegna} Figure 1 shows the
Zipf plot for noise with various correlation exponents, $\beta$,  for
$6$-tuples from 72k data point sequences.  The data cleanly fit a
power-law over
about three decades, similar to or better than the non-coding DNA
results of ref.~1.  A monotonic increase in $\zeta$ with increasing $\beta$ is
observed as shown in the inset.  The power-law scaling breaks down for
$\beta\approx 1$ or larger but is recovered when a larger alphabet is
used.~\cite{ref:tobe} Also shown are the redundancy percentages, $R$,
for the noise,
calculated as in ref.~1.  The $R$ values are similar to those in ref.~1
with similar z values.

The best fit to a power-law in a Zipf plot in ref.~1
(fig.~1) is from mammalian DNA  with $\zeta=0.289$.  The correlation
exponents found for certain mammalian primarily non-coding DNA were in
the range $0.64 < a < 0.71$ which give $0.28 < b < 0.42$.~\cite{ref:peng} By
comparison we find $\zeta=0.28$ for noise with $\beta=0.30$. This together with
the fact that {\em coding} DNA has $\beta\approx\zeta\approx 0$ are
consistent with the
idea that power-law correlations without a linguistic component could
account for the behavior reported in ref.~1.   Our results demonstrate that the
Zipf power-law scaling and non-zero Shannon redundancies alone must
not be relied upon to distinguish language from noise.  However, a
detailed comparison of Zipf exponents for DNA, language, and noise
with the same correlation exponent might be revealing.  A full account
of these findings will appear elsewhere.~\cite{ref:tobe}

We thank W-J. Rappel, and T. Sage for helpful discussions.  This work
was supported by NSF/DMR-9458008 (NYI) and NSF/ECS-9102396.

\section*{Figure Caption}
Zipf plot, exponents, and redundancy \% for 6-tuples from power-law noise

\end{document}